\documentclass[12pt,preprint]{aastex}
\usepackage{amsmath}
\usepackage{graphicx}
\usepackage{multicol}
\usepackage{float}
\usepackage{amsmath,esint}

\shorttitle{Revisiting VHE Gamma-Ray Absorption in Cosmic Propagation under the Combined effects of ALPs and LIV}
\shortauthors{LH Qin et al.}

\begin{document}


\title{Revisiting Very High Energy Gamma-Ray Absorption in Cosmic Propagation under the Combined Effects of Axion-Like Particles and Lorentz Invariance Violation}

\author{Longhua Qin\altaffilmark{1}, Jiancheng Wang\altaffilmark{2,3},Chuyuan Yang\altaffilmark{2,3,*}, Huaizhen Li\altaffilmark{1,*}, Quangui Gao\altaffilmark{1,*}, Ju Ma\altaffilmark{1}, Ao Wang\altaffilmark{1}, Weiwei Na\altaffilmark{1}, Ming Zhou\altaffilmark{2,3}, Zunli Yuan\altaffilmark{4}, Chunxia Gu\altaffilmark{5} and Guangbo Long\altaffilmark{6}}

\affil{1. Department of Physics, Yuxi Normal University, Yuxi,Yunnan, 653100, People’s Republic of China}
\affil{2. Yunnan Observatory, Chinese Academy of Sciences, Kunming, Yunnan, 650011, People’s Republic of China}
\affil{3. Key Laboratory for the Structure and Evolution of Celestial Objects, Chinese Academy of Sciences, Kunming, Yunnan, 650011, People’s Republic of China}
\affil{4. Department of Physics and Synergistic Innovation Center for Quantum Effects and Applications, Hunan Normal University, Changsha,Hunan, 410081, People’s Republic of China}
\affil{5. Oxbridge College, Kunming University of Science and Technology, Kunming, Yunnan, 650011, People’s Republic of China}
\affil{6. School of Intelligent Engineering, Shaoguan University, Shaoguan, Guangdong, 512005, People’s Republic of China}
\affil{* Corresponding author:lhz@yxnu.edu.cn;qggao@yxnu.edu.cn;chyy@ynao.ac.cn}

\clearpage

\begin{abstract}
Very-high-energy (VHE; $E \gtrsim 100$ GeV) gamma rays are expected to experience strong attenuation during cosmological propagation due to electron-positron pair production on the extragalactic background light (EBL). Recent observations of GRB 221009A (z = 0.151), including photons up to $\sim 18$ detected by LHAASO and a $\sim 300\ \mathrm{TeV}$ event reported by Carpet-3, suggest a higher-than-expected transparency of the Universe at extreme energies. These observations cannot be explained by standard EBL absorption alone; moreover, neither Lorentz invariance violation (LIV) nor photon-axion-like particle (ALP) oscillations, when considered in isolation, appear sufficient to account for the survival of such photons over cosmological distances. In this work, we propose a joint propagation scenario that incorporates photon-ALP mixing in astrophysical magnetic fields together with subluminal quadratic LIV corrections to the $\gamma\gamma$ pair-production threshold. Applying this framework to the broadband gamma-ray spectrum of GRB 221009A, we show that ALPs with coupling ($g_{a\gamma} = 1.685 \times 10^{-10}\mathrm{GeV}^{-1}$ ) and mass ($m_a = 9.545 \times 10^{-8}\mathrm{eV}$), combined with a quadratic LIV energy scale ($E_{\rm LIV,2} = 1.30 \times 10^{-7} E_{\rm Pl}$) adopted from the literature, can significantly enhance the photon survival probability in the energy range (10\text{-}300) TeV. The resulting enhancement exceeds that obtained from either ALP mixing or LIV effects alone. These results indicate that a combined ALP-LIV scenario may provide a viable interpretation of the extreme-energy gamma-ray observations of GRB 221009A and highlight the potential of VHE gamma-ray measurements as probes of physics beyond the Standard Model.
\end{abstract}


\clearpage

\section{Introduction} \label{1. Introduction}

Very-high-energy (VHE) gamma rays, typically defined as photons with energies exceeding 100 GeV, are expected to undergo substantial attenuation as they propagate through the intergalactic medium. This attenuation is primarily caused by interactions with the extragalactic background light (EBL), which lead to electron-positron pair production ($\gamma\gamma \rightarrow e^{+}e^{-}$). Beyond this primary absorption process, the resulting $e^{+}e^{-}$ pairs can further up-scatter cosmic microwave background (CMB) photons via inverse-Compton interactions, thereby initiating electromagnetic cascades that regenerate secondary gamma rays at lower energies. Together, these processes can significantly reshape the observed spectra of high-energy gamma-ray sources. Moreover, due to the strong EBL absorption, gamma rays with energies above $\sim$30 TeV are expected to have extremely limited propagation distances on cosmological scales, making their detection particularly challenging \citep[see e.g.,][]{qinl2023,dom2013}.

Despite theoretical expectations of substantial attenuation due to pair production with EBL, observations of VHE gamma rays from distant blazars consistently reveal a level of transparency exceeding standard model predictions \citep[see e.g.,][]{car2024}. This tension poses a significant challenge to our understanding of gamma-ray propagation. Initial evidence arose from H.E.S.S. observations of blazars such as H 2356-309 ($z = 0.165$) and 1ES 1101-232 ($z = 0.186$), whose spectra could only be reconciled with EBL densities substantially below established models \citep{aha2006}. Additional support came from MAGIC observations of the high-redshift AGN 3C 279 ($z = 0.54$) \citep{mag2008}. A systematic trend of reduced absorption in the VHE spectral slopes of AGN at $z > 0.2$ was also reported by \cite{dea2013}, further reinforcing the anomaly. Besides, the recent detection of multi-TeV gamma rays from GRB 221009A, including photons up to 18 TeV observed by LHAASO \citep{caoz2023} and a 251 TeV photon candidate reported by the Carpet collaboration \citep{dzh2022} later updated as 300$^{-30}_{+42}$ TeV by  \citep{dzh2025}, presents a significant tension with standard models of gamma-ray propagation. Under conventional EBL attenuation models, the observed spectrum-associated with the most luminous GRB ever recorded-would necessitate an unphysical intrinsic spectral upturn to account for the transmitted flux at these energies, even when adopting the lowest EBL density estimates. This discrepancy strongly motivates the investigation of beyond-standard-model scenarios, such as axion-like particles (ALPs) and Lorentz invariance violation (LIV), that could effectively reduce EBL-induced attenuation and increase the gamma-ray transparency of the universe.

ALPs and LIV provide distinct frameworks to explain the anomalous transparency of VHE gamma rays. ALPs introduce a new light pseudo-scalar particle that can oscillate with photons in cosmic magnetic fields, allowing gamma rays to partially evade attenuation by EBL. Such ALPs need to convert back to photons in the vicinity of our observatories, e.g., in the galactic magnetic field. In addition, LIV scenarios may introduce a preferred reference frame, allow otherwise forbidden processes (see e.g., photon decays or vacuum Cherenkov radiation; \citealt{leh2004, sav2024}), and modify fundamental interaction cross-sections \citep[e.g.,][]{car2024}. These effects can lead to additional energy losses in high-energy electromagnetic cascades, thereby affecting the propagation of high-energy gamma-ray photons in the intergalactic medium. Historically, several studies have shown that photon-ALPs mixing can enhance the transparency of the Universe to VHE gamma rays \citep[e.g.,][]{dea2008, lon2020, caoz2023, chen2024}, while interest in LIV-induced modifications to the pair-production threshold grew following observations of high energy blazars \citep[e.g.,][]{kif1999, ste2001} and, more recently, gamma ray bursts such as GRB 221009A \citep[e.g.,][]{fin2023, zhe2023}. Searches for ALP-induced spectral irregularities have constrained ALP-photon couplings across wide mass ranges  \citep[see e.g.,][]{aje2016}, and observations of rapid TeV variability in gamma-ray bursts have set stringent limits on LIV energy scales \citep[see e.g.,][]{acc2020}. However, there is no fundamental reason to exclude the simultaneous presence of ALPs and LIV in explaining the anomalous transparency of gamma rays over cosmological distances \citep{gal2025}. While photon-LP mixing \citep[e.g.,][]{dea2008,lon2020,caoz2023,chen2024} and subluminal Lorentz invariance violation (LIV)-induced modifications to the pair-production threshold \citep[e.g.,][]{kif1999,ste2001,fin2023,zhe2023} have each been extensively explored as possible explanations for anomalies in VHE gamma-ray transparency-including recent applications to the LHAASO and Carpet-3 observations of GRB 221009A-neither mechanism, when considered in isolation, is able to reproduce the broadband spectrum over the full 10–300 TeV range without invoking extreme parameter choices or encountering tension with independent constraints. These include air-shower limits on LIV energy scales \citep[e.g.,][]{gal2008,ell2019} and bounds on ALP couplings inferred from spectral irregularities and other astrophysical probes \citep[e.g.,][]{hor2012,acc2020}. Recent quantitative analyses of the combined LHAASO-Carpet-3 fluence spectrum by \citet{sat2025} indicate that ALP-induced effects can provide an improved statistical description compared to LIV-only scenarios, although the two mechanisms are still treated independently.

Based on the above considerations, we construct a model of TeV gamma-ray absorption that simultaneously accounts for both ALPs and LIV, and evaluate its viability using observational data from GRB 221009A. As the most energetic gamma-ray burst ever recorded (redshift $z = 0.151$) \citep{deu2022}. GRB 221009A exhibits exceptional brightness and broad spectral coverage \citep{caoz2019,caoz2023}, offering a unique opportunity to probe $\gamma\gamma$ attenuation over cosmological distances. More than $6\times10^{4}$ gamma-ray events with energies between 200~GeV and 7~TeV were detected by the LHAASO Water Cherenkov Detector Array (WCDA) within 3000~s after the Fermi-GBM trigger \citep{caoz2023}. LHAASO also reported the first detection of photons with energies exceeding 10~TeV from a gamma-ray burst, GRB 221009A \citep{caoz2023}. The Kilometer Square Array (KM2A) recorded 142 photon-like events in the 3–20~TeV range during 230-900~s after the trigger, compared to an expected background of 16.7 events, among which nine events have reconstructed energies $\gtrsim 10$~TeV and background probabilities between 0.045 and 0.17. In addition, the Carpet-3 air shower array at the Baksan Neutrino Observatory reported the detection of an air-shower event spatially and temporally coincident with GRB~221009A, with an estimated primary energy of $300^{+43}_{-38}$~TeV \citep{dzh2025}.

We first compute the photon survival probability, $P_{\gamma \rightarrow \gamma}(E, z)$, for VHE gamma rays under three scenarios: ALP-only, LIV-only, and the combined ALP-LIV framework. Using the survival probabilities derived in each case, we then perform spectral fits to the multi-band observations of the GRB. For the ALP-LIV case, the fitting is carried out within the simplified hybrid model described above.

This paper is organized as follows. Section~\ref{mt} outlines the physical processes relevant to the propagation of high-energy gamma rays. Specifically, we discuss: (I) the prompt gamma-ray emission and attenuation due to interactions with the EBL in Section~\ref{ebl}; (II) ALP oscillations in external magnetic fields in Section~\ref{alp}; (III) the impact of LIV on the $\gamma\gamma \rightarrow e^{+}e^{-}$ pair-production process in Section~\ref{stlv}; and (IV) the combined effects of ALPs and LIV, together with the corresponding model simplifications adopted in this work, in Section~\ref{alplv}. Finally, Section~\ref{dis} summarizes our results and presents the discussion and conclusions.

\section{Model and strategy}\label{mt}

\subsection{Prompt gamma-ray spectrum and absorption on EBL}\label{ebl}

To reconstruct the intrinsic emission spectrum, we adopt a power-law form,$\phi_{\rm PL}(E) = \phi_{0}(E/E_{0})^{-\alpha},$
where $\phi_0$ is the normalization constant and $E_0 = 1\ \mathrm{TeV}$ is the reference energy and $\alpha$ denotes the spectral index. The resulting observed spectral energy distribution (SED) is then given by
\begin{equation}
\label{SED}
E^2\frac{{\rm d} N}{{\rm d} E} = P_{\gamma \rightarrow \gamma}(E, z) \phi_{\rm PL}(E),
\end{equation}
in units of erg cm$^{-2}$ s$^{-1}$. 

The optical depth $\tau(E, z)$ for a gamma-ray photon with observed dimensionless energy $\epsilon_s = E / (m_e c^2)$, interacting with background photons of proper-frame energy density $u_p(\epsilon_p; z)$, is given by \cite{gou1967}

\begin{flalign}
\label{taugg_model}
\tau(E, z) & = \frac{c\pi r_e^2}{\epsilon_s^2 m_ec^2}\ 
\int^z_0 \frac{dz^\prime}{(1+z^\prime)^2}\ 
\left| \frac{dt_*}{dz^\prime}\right|\ 
\nonumber \\ & \times
\int^\infty_{\frac{1}{\epsilon_s(1+z^\prime)}} d\epsilon_{p} 
\frac{ \epsilon_{p}u_{p}(\epsilon_{p};z^\prime)}
{\epsilon_p^{4}}\ 
\bar{\phi}(\epsilon_{p}\epsilon_s(1+z^\prime))\ ,
\end{flalign}

where $r_e$ is the classical electron radius, $m_e$ is the electron mass, and $\bar{\phi}(s_0)$ encodes the pair-production cross-section as a function of the center-of-mass energy squared \citep[see e.g.,][]{gou1967}. The differential time-redshift relation in a standard flat $\Lambda$CDM cosmology is:

\begin{equation}
\frac{dt}{dz} = \frac{-1}{H_0 (1+z)\sqrt{\Omega_m(1+z)^3 + \Omega_\Lambda}},
\end{equation}

here, a flat $\Lambda$CDM cosmology is adopted, where $\Omega_{\Lambda} = 0.7$, $\Omega_M = 0.3$, and $H_0 = 70\rm {km\ s^{-1}\ Mpc^{-1}}$. Additionally, the unprimed and primed quantities denote the quantities in the observer's frame and the co-moving frame, respectively.

\subsection{ALP-photon oscillation in external magnetic fields}\label{alp}

ALPs are pseudo-scalar bosons predicted in various extensions of the Standard Model, particularly in string theory and extra-dimensional frameworks. Unlike the Quantum Chromodynamics (QCD) axion \citep[see e.g.,][]{pec1977}, ALPs are not required to solve the strong CP problem and can span a wide parameter space in mass $m_a$ and photon coupling $g_{a\gamma}$ \citep[see e.g.,][]{abb1983,con2006}. Their interaction with photons is described by the Lagrangian:

\begin{equation}\label{ll}
\mathcal{L}_{\rm int} = -\frac{1}{4}g_{a\gamma} a F_{\mu\nu} \tilde{F}^{\mu\nu} = g_{a\gamma} a\, \mathbf{E} \cdot \mathbf{B},
\end{equation}

where $a$ is the ALPs field, $F_{\mu\nu}$ the electromagnetic field tensor, and $\tilde{F}^{\mu\nu}$ its dual. This coupling enables photon-ALP conversions in external magnetic fields \citep{raf1988}, a process that can significantly affect high energy photon propagation over cosmological distances.

The evolution of an photon-ALP beam propagating through an external magnetic field along the $z$-axis is governed by a Schrödinger-like equation that accounts for mixing between photons and ALPs. The system is described by the state vector $\psi(z) = (A_x, A_y, a)^T$, 

\begin{equation}
i \frac{d\psi(z)}{dz} = M \psi(z),
\end{equation}
where $A_x$ and $A_y$ denote the photon amplitudes for linear $x$- and $y$-polarizations, and $a$ is the ALPs amplitude. The mixing matrix $M$, expressed in the interaction basis\citep{lon2020}, is given by:

\begin{equation}\label{eq:MM}
M = 
\begin{pmatrix}
\Delta_\perp & 0 & 0 \\
0 & \Delta_\parallel & \Delta_{a\gamma} \\
0 & \Delta_{a\gamma} & \Delta_{aa}
\end{pmatrix},
\end{equation}
where $ \Delta_\perp = \Delta_{\text{pl}} + 2\Delta_{\text{QED}} + \Delta_{\text{CMB}} + \frac{i}{2\lambda_\gamma}$, and 
$\Delta_\parallel = \Delta_{\text{pl}} + \frac{7}{2} \Delta_{\text{QED}} + \Delta_{\text{CMB}} + \frac{i}{2\lambda_\gamma}$ represent the effective dispersion relations for the $x$- and $y$-polarized photon modes, including contributions from plasma effects, vacuum birefringence due to quantum electrodynamics (QED), photon scattering \citep{dob2015} with the cosmic microwave background (CMB), and absorption by EBL \citep[e.g.,][]{lon2020}. Noted from Eq.~(\ref{ll}) that ALPs couple only to the photon polarization component lying in the plane defined by the transverse magnetic field $B_T$ and the propagation direction. By choosing a coordinate frame in which $B_T$ is aligned along the $y$-axis, only the $A_y$ component mixes with the ALPs, thereby simplifying the dynamics. The off-diagonal term $\Delta_{a\gamma} = \frac{1}{2} g_{a\gamma} B_T$ quantifies the photon–ALPs coupling strength, proportional to the photon–ALPs coupling constant $g_{a\gamma}$ and the transverse magnetic field $B_T$. The ALPs diagonal term $\Delta_{aa} = -\frac{m_a^2}{2 E_\gamma}$ accounts for the effective mass contribution of the ALPs, where $E_\gamma$ is the photon energy. The numerical values for the different $\Delta$ can be found in \cite{hor2012}:

\begin{equation}
\Delta_{a\gamma} \simeq 1.52 \times 10^{1} \left(\frac{g_{a\gamma}}{10^{-11} \, {\rm GeV}^{-1}} \right)
\left(\frac{B_T}{10^{-9} \, \rm G}\right) \, {\rm kpc}^{-1},
\end{equation}

\begin{equation}
\Delta_a \simeq -7.8 \times 10^{-3} \left(\frac{m_a}{10^{-3} \, {\rm TeV}}\right)^2 \left(\frac{E}{10 \, {\rm eV}} \right)^{-1} \, {\rm kpc}^{-1},
\end{equation}

\begin{equation}
\Delta_{\rm pl} \simeq - 1.1\times 10^{-10} \left(\frac{E}{{\rm TeV}}\right)^{-1} \left(\frac{n_e}{10^{-3} \, {\rm cm}^{3}}\right) \, {\rm kpc}^{-1},
\end{equation}

where $n_e$ denotes the electron number density. The plasma term $\Delta_{\rm pl}$ is primarily relevant in relatively dense astrophysical environments, such as the host galaxy or the Milky Way disk. In the intergalactic medium, its contribution becomes negligible due to the extremely low electron density.

The system’s statistical evolution is captured by the density matrix $\rho(z) = \psi(z) \psi^\dagger(z)$, which obeys the von Neumann-like equation \citep{bas2010}:

\begin{equation}
i \frac{d\rho(z)}{dz} = [\rho(z), M],
\end{equation}

ensuring the conservation of probability and Hermiticity. In a homogeneous region of size $\Delta z$, the formal solution is \citep[e.g.,][]{hor2012,dea2013}:

\begin{equation}
\psi(z + \Delta z) = U \psi(z), \quad \text{with} \quad U = \exp(-i M \Delta z),
\end{equation}

\begin{equation}
\rho(z + \Delta z) = U \rho(z) U^\dagger,
\end{equation}

where $U$ is the unitary evolution operator over the interval $\Delta z$.

For an initially unpolarized photon beam with no ALPs component, the initial density matrix is:

\begin{equation}
\rho(0) = \frac{1}{2}
\begin{pmatrix}
1 & 0 & 0 \\
0 & 1 & 0 \\
0 & 0 & 0
\end{pmatrix}.
\end{equation}

The probability of a photon remaining a photon after propagation is obtained by tracing over the photon subspace \citep[e.g.,][]{hor2012,dea2013,lon2020}:

\begin{equation}
P_{\gamma \rightarrow \gamma}(z) = \mathrm{Tr} \left[
\begin{pmatrix}
1 & 0 & 0 \\
0 & 1 & 0 \\
0 & 0 & 0
\end{pmatrix}
\rho(z)
\right] = \rho_{11}(z) + \rho_{22}(z),
\end{equation}

while the photon-to-ALPs conversion probability is $P_{\gamma \rightarrow a}(z) = \rho_{33}(z)$.

\subsection{Effects of LIV on pair production process} \label{stlv}
Lorentz invariance, a cornerstone of special relativity, may be violated in some quantum gravity frameworks such as string theory and loop quantum gravity \citep[see e.g.,][]{ame1998, mat2005, ell2008}. In such LIV scenarios, the photon dispersion relation is modified as

\begin{equation}
\label{livdr}
E^2 - p^2c^2 = \pm E^2 \left(\frac{E}{E_{\rm QG}}\right)^n,
\end{equation}

where $E_{\rm QG}$ is the quantum gravity scale (near $E_{\rm Planck} \sim 10^{28}~\mathrm{eV}$), and $n$ denotes the leading-order correction,that $n = 1$ or $2$ corresponds to linear or quadratic leading-order LIV corrections, respectively. The ''$+$'' represents superluminal LIV and the ''$-$'' sign corresponds to the subluminal case. LIV leads to two notable astrophysical consequences. First, photon speed is energy-dependent, resulting in an energy-dependent time delay for photons emitted simultaneously from distant transients such as GRBs. Time-of-flight measurements from such sources have placed lower bounds on $E_{\rm QG}$ \citep[see e.g.][]{abd2009,ell2019}. Second, and more relevant to gamma ray absorption, LIV exerts substantial impacts on the propagation of TeV-PeV gamma rays in intergalactic space by modifying interaction kinematics and enabling novel processes forbidden in the Standard Model. For electromagnetic processes relevant to gamma rays, superluminal LIV scenarios permit photon decay in vacuum, and the detection of PeV-energy gamma rays has placed stringent constraints on the LIV energy scale; in contrast, subluminal LIV can elevate the energy threshold of the $\gamma + \gamma \rightarrow e^+ + e^-$ interaction, enhancing the Universe’s transparency to gamma rays above ~10 TeV and potentially enabling the detection of more distant sources. LIV also modifies the threshold energies for interactions like pair production and inverse Compton scattering, altering the development of electromagnetic cascades-such modifications are more pronounced in LIV frameworks. Additionally, LIV-induced changes to cross sections and decay rates, though less explored, may further reshape the gamma-ray flux spectrum at Earth, while the interplay between LIV and background radiation fields (CMB and EBL) influences interaction rates and propagation lengths.

The photon energy in the threshold condition for pair production is modified. This is implemented by replacing $\epsilon_s$ with an effective energy \citep[e.g.,][]{fin2023}:

\begin{equation}
\label{taugg_LIV}
\epsilon_s \rightarrow \frac{\epsilon_s}
{1 + \frac{1}{4} \left( \frac{\epsilon_s m_e c^2}{E_{\rm QG}} \right)^n \epsilon_s^2 }.
\end{equation}

In this work, we implement the LIV correction solely through this threshold modification (Eq.~(\ref{taugg_LIV})). Although a complete LIV treatment would also include modifications to the squared interaction amplitude $|M|^2$, the full cross section beyond the simple threshold shift, and potential vacuum processes (e.g., photon decay or vacuum Cherenkov radiation, \citet{car2024}), we focus exclusively on the threshold correction. This choice is motivated by the fact that, in the relevant energy range (TeV-PeV), the threshold shift overwhelmingly dominates the suppression of the optical depth $\tau_{\gamma\gamma}$, while additional amplitude and cross-section changes are subdominant and do not qualitatively alter the transparency enhancement \citep{car2024, jac2008}. Thus, focusing on the threshold modification captures the leading-order LIV effect on gamma-ray propagation while maintaining computational tractability and model simplicity.

\subsection{Joint Effects of ALP-Lorentz Invariance Violation and Model Simplification}\label{alplv}

Gamma-ray photons originating from TeV sources may undergo continuous interconversion with ALPs as they propagate through various astrophysical environments, including the jet and host galaxy, the intergalactic medium, and the Milky Way. In this process, the potential influence of LIV needs to be considered. When the combined effects of ALPs and LIV are treated in a unified framework, the ALP contribution is fully incorporated through the evolution of the mixing matrix $M$ (Eq.~(\ref{eq:MM})), with the potential LIV effect manifesting itself in this matrix throughout the propagation process.

The LIV contribution, arising from the modified photon dispersion relation in Eq.~(\ref{livdr}) under the subluminal case, alters the kinematics of $\gamma\gamma$ pair production. Within a complete effective field theory framework, this modification leads to two primary effects. Firstly, LIV induces a substantial upward shift in the pair-production threshold. To quantify this effect, we evaluate the gamma-ray mean free path $\lambda_\gamma$ using an expression for the optical depth $\tau(E, z)$ that explicitly accounts for LIV corrections. The mean free path of a photon with energy $E$ emitted at redshift $z$ is given by \citep[e.g.,][]{lon2020}
\begin{equation}
\lambda_\gamma(E, z) = \frac{D(z)}{\tau(E, z)},
\end{equation}
where $D(z)$ denotes the comoving distance to the source. The LIV-induced threshold shift is implemented by replacing the soft-photon energy $\epsilon_s$ in Eq.~(\ref{taugg_model}) with the effective value defined in Eq.~(\ref{taugg_LIV}) \citep[see, e.g.,][]{car2024}. Secondly, LIV introduces an additional energy-dependent correction term,
\begin{equation}
\Delta_{\rm LIV} \simeq -\frac{1}{2E}\left(\frac{E}{E_{\rm QG}}\right)^2,
\end{equation}
which enters directly into the diagonal photon terms $\Delta_\perp$ and $\Delta_\parallel$ of the mixing matrix $M$. This correction modifies the photon phase velocities and can further influence the oscillation phases and probabilities of the photon--ALP system during propagation \citep[see, e.g.,][]{gal2008}.


Gamma-ray photons emitted from a TeV source propagate through the magnetic fields of the jet/host system (S), intergalactic space (I), and the Milky Way (M) before reaching the observer. The general expression for the photon survival probability is:

\begin{align}
P_{\gamma \rightarrow \gamma} =\ 
& P_{\gamma \rightarrow \gamma}^{\text{S}} \left(
    P_{\gamma \rightarrow \gamma}^{\text{I}} P_{\gamma \rightarrow \gamma}^{\text{M}}
  + P_{\gamma \rightarrow a}^{\text{I}} P_{a \rightarrow \gamma}^{\text{M}} 
\right) \nonumber \\
+\
& P_{\gamma \rightarrow a}^{\text{S}} \left(
    P_{a \rightarrow \gamma}^{\text{I}} P_{\gamma \rightarrow \gamma}^{\text{M}}
  + P_{a \rightarrow a}^{\text{I}} P_{a \rightarrow \gamma}^{\text{M}} 
\right).
\end{align}

We neglect photon-ALP mixing in the intergalactic magnetic field (IGMF), such that
\begin{equation}
P^{\rm I}_{\gamma\to\gamma} = e^{-\tau_{\gamma\gamma}}, \quad
P^{\rm I}_{\gamma\to a} = P^{\rm I}_{a\to\gamma} = 0.
\end{equation}

This approximation is justified by the fact that, for typical IGMF strengths $B_{\rm IGMF} \lesssim 1$ nG and a source redshift $z = 0.151$, the photon-ALP conversion probability remains negligible at the highest energies considered here ($E \gtrsim 10$ TeV). The dominant suppression arises from the weak magnetic field and dephasing effects over cosmological distances, with plasma contributions being subdominant due to the extremely low electron density ($n_e \sim 10^{-7}$ cm$^{-3}$ or lower in voids) \citep{dur2013, alv2021}. Although the long propagation path length could partially compensate for the small $B_{\rm IGMF}$ (as encoded in the oscillation probability for large-scale low-B regions, Eq.~\ref{eq:extragalactic_P}), the net conversion remains negligible in this redshift and energy regime \citep[see, e.g.,][]{lon2021, chen2024}. We further assume negligible photon-to-ALP conversion in the Milky Way, i.e., $P_{\gamma \rightarrow \gamma}^{\text{M}} \approx 1$, consistent with typical Galactic electron densities and magnetic field geometries along the line of sight to GRB 221009A \citep[see, e.g.,][]{lon2021, chen2024}. For the source region, we approximate the photon survival probability as $P_{\gamma \rightarrow \gamma}^{\text{S}} \approx 1 - P_{\gamma \rightarrow a}^{\text{S}}$. As discussed in Section~\ref{stlv}, our treatment of LIV within the ALP framework is intentionally restricted to the modification of the pair-production energy threshold, which enters through the photon mean free path in the propagation matrix $M$. Within this simplified but well-motivated approach, LIV effects are incorporated solely via the replacement $\tau_{\gamma\gamma} \rightarrow \tau_{\mathrm{LIV},\gamma\gamma}$, as defined in Eq.~(\ref{taugg_LIV}), with all other contributions kept fixed. Under these assumptions, the total photon survival probability simplifies to $P_{\gamma \rightarrow \gamma} \approx (1 - P_{\gamma \rightarrow a}^{\text{S}}) e^{-\tau_{\text{LIV},\gamma \gamma}} + P_{\gamma \rightarrow a}^{\text{S}} P_{a \rightarrow \gamma}^{\text{M}}.$ Photon-ALP oscillations arise in transverse magnetic fields, with efficiencies set by the competition between the mixing term and other momentum-mismatch contributions. In the strong-mixing regime-realized for sufficiently small ($m_a$) and/or strong magnetic fields-the conversion becomes efficient and approximately energy independent, the relation $P_{a \rightarrow \gamma} = 2 P_{\gamma \rightarrow a}$ holds, as $P_{\gamma \rightarrow a} \to 1/3$ \citep[see e.g.][]{lon2020,chen2024}. In the single domain approximation, the photon conversion probability is given by \cite{mir2009,dea2011,gal2018}:

\begin{equation}\label{eq:extragalactic_P}
P_{\gamma\rightarrow a}(E) = \sin^2(2\theta)\sin^2\left(\frac{\Delta_{\rm osc} L}{2}\right),
\end{equation}
where
\begin{align}
\tan(2\theta) &= \frac{2\Delta_{a\gamma}}{\Delta_{aa} - \Delta_\parallel}, \\
\Delta_{\rm osc} &= \sqrt{(\Delta_{aa} - \Delta_\parallel)^2 + 4\Delta_{a\gamma}^2}.
\end{align}

To simplify the modeling of photon-ALP mixing, we neglect the QED vacuum polarization and the CMB-induced photon dispersion terms in the mixing matrix. This is justified by the approximate scaling relations $\Delta_{\mathrm{QED}} \propto E_\gamma \cdot B_T^2$ and $\Delta_{\mathrm{CMB}} \propto E_\gamma \cdot (1 + z)^3$. Given that the transverse magnetic field $B_T$ is of the order of a few microgauss and the redshift of the GRB is $z < 0.5$, similar to the case considered in \citep{chen2024}, both contributions are subdominant compared to the photon-ALP coupling and plasma effects in the energy range of interest.

Additionally, we account for the stochastic orientation of the magnetic field in each domain, as considered in \citep{gro2002,chen2024}. When the overall magnetic region has a size $L \gg r$, where $r$ is the coherence length of a single domain and $P_0$ is the photon-to-ALPs conversion probability within a single domain, the expression for the total conversion probability was originally derived in \citep{gro2002} and is adopted in the present work following the formulation provided by \citep{chen2024}.

To model the propagation of VHE gamma rays within the source environment, we adopt representative physical parameters for the GRB host galaxy. Specifically, we assume a transverse magnetic field strength of $B_T \simeq 0.5\,\mu{\rm G}$, an electron density of $n_e \simeq 0.04\,{\rm cm^{-3}}$, and a characteristic propagation length of $\sim 10\,{\rm kpc}$ \citep{fle2010}. These values are consistent with typical conditions in disk regions of GRB host galaxies and are commonly adopted in similar analyses (e.g., see \citealt{chen2024}).

For photon-ALP reconversion in the Milky Way, we follow the Galactic magnetic field model of \citet{jan2012} and the electron density distribution of \citet{cor2002}, adopting a regular magnetic field strength of $\sim 3.0\,\mu{\rm G}$, an electron density of $n_e \simeq 0.05\,{\rm cm^{-3}}$, and a propagation distance of $\sim 30\,{\rm kpc}$ along the line of sight to GRB~221009A (Galactic coordinates $l \simeq 22.4^\circ$, $b \simeq -42.5^\circ$). These parameters are representative of typical Galactic disk conditions. Moreover, the turbulent magnetic field component along this direction is expected to have a limited impact in the strong-mixing regime relevant to our study \citep{ung2024}.

\section{Application to GRB 221009A}\label{appgrb}

We utilized the observational data of GRB 221009A from LHAASO \citep{caoz2023}, specifically the measurements from the Kilometer Squared Array (KM2A) and the Water Cherenkov Detector Array (WCDA) during the 300-900 s interval after the trigger. In addition, the Carpet-3 experiment detected gamma-ray photons with energies up to 300 TeV \citep{dzh2025}. Since the differential flux values at 300 TeV are not explicitly provided in \citep{dzh2025}, we estimated the upper and lower limits of the 300 TeV flux by assuming time intervals of one hour and one day for temporal differentiation, along with the corresponding average value. This procedure follows the note that the burst duration is poorly constrained, ranging from roughly one hour to an entire day and that the flux can be inferred by dividing the reported differential fluence by the adopted time window.

We perform Bayesian parameter inference using the Python package UltraNest \citep{buc2021} to obtain posterior distributions of model parameters. The LIV-modified EBL optical depth is calculated using the publicly available \texttt{ebltable} package\footnote{\url{https://github.com/me-manu/ebltable/blob/main/CITATION.cff}} \citep{mey2022}, with the EBL model of \citet{fin2022} adopted throughout this work.

We assess the performance of different photon survival probability models in fitting the observed spectrum of GRB~221009A using the Akaike Information Criterion (AIC), which balances goodness-of-fit against model complexity and penalizes excessive free parameters \citep[see e.g.,][]{aka1974}. For each scenario, the maximum likelihood $\hat{L}$ is obtained and the AIC is computed as $\mathrm{AIC} = 2k_{n} - 2\ln(\hat{L})$, where $k_{n}$ denotes the number of free parameters. The model with the minimum AIC value is identified as the preferred description of the data, providing a quantitative basis for comparing alternative photon survival prescriptions.

\begin{figure}[h!t]
    \centering
    \includegraphics[width=5.0in, angle=0]{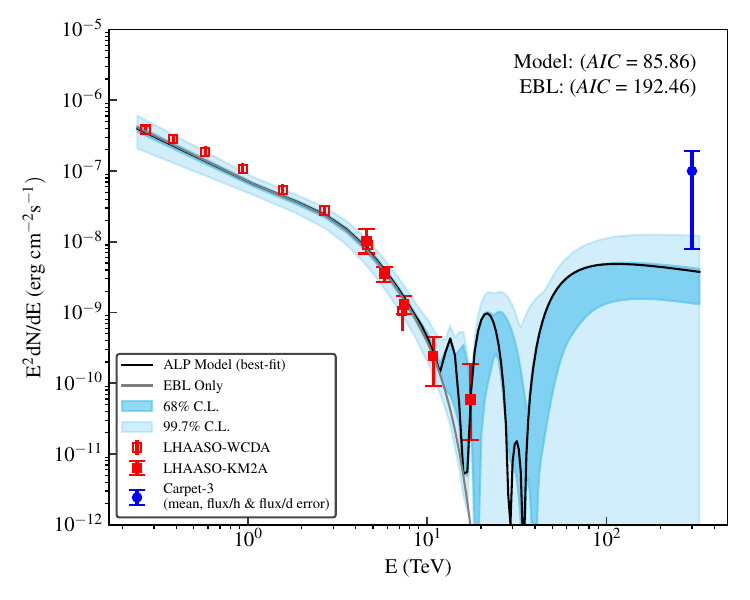}  
    \caption{Corresponding best-fit SEDs of GRB 221009A under only ALPs scenario with a coupling constant $ g_{a\gamma} = 1.489 \times 10^{-10} \, \mathrm{GeV}^{-1} $ and $m_a = 4.191 \times 10^{-7} \, \mathrm{eV} $. The Carpet-3 point at ~300 TeV is derived by dividing the reported differential fluence by time windows of 1 hour (upper bound) and 1 day (lower bound), with the mean shown.}
    \label{fig:OnlyALPs} 
\end{figure}

\begin{figure*}[t!]
    \centering
    \includegraphics[width=\textwidth]{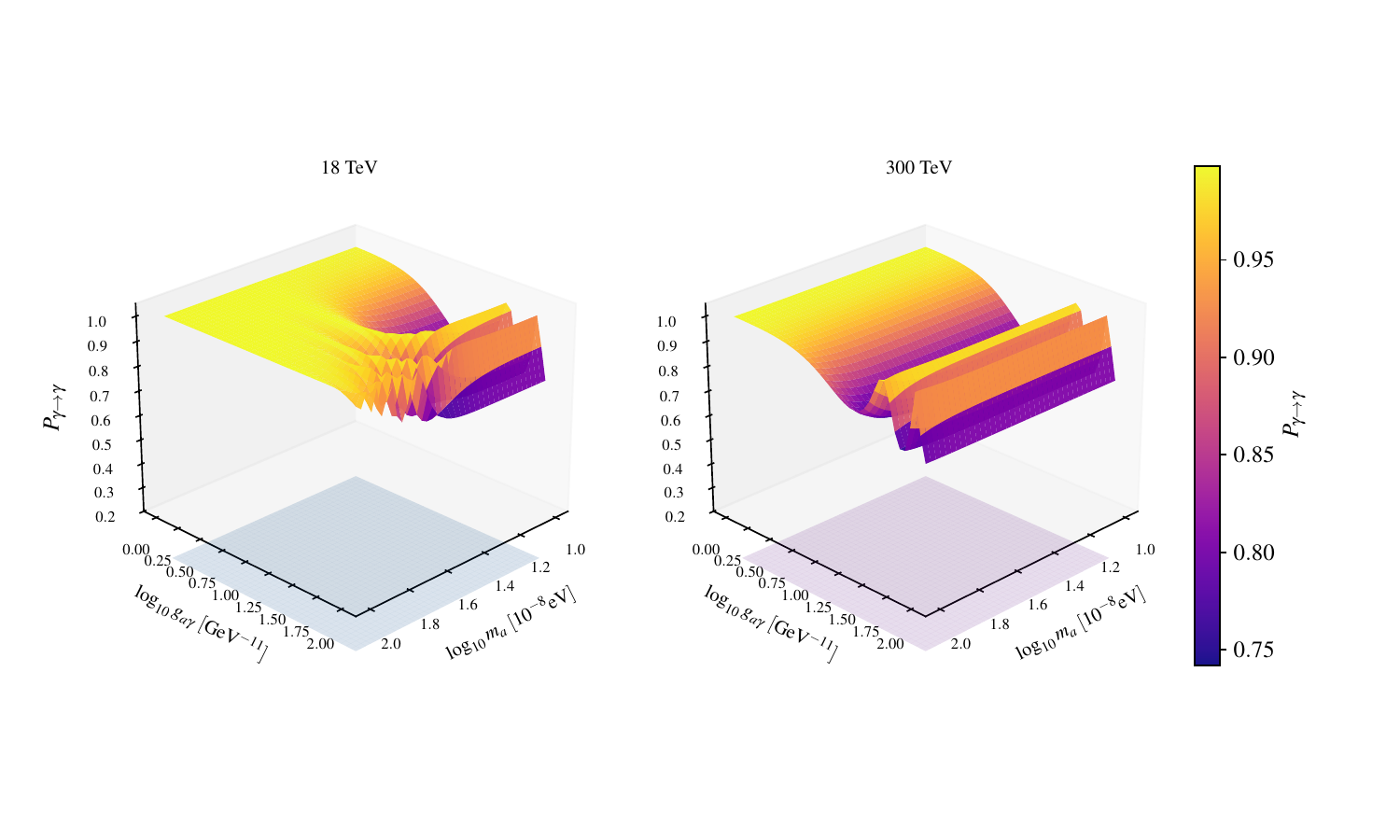}  
    \caption{Photon survival probability within the host galaxy of GRB~221009A as a function of ALPs mass ($m_a$) and the ALP-photon coupling constant ($g_{a\gamma}$), evaluated for photon energies of 18~TeV and 300~TeV. The results indicate that a substantial fraction of photons can survive across the explored $m_a$–$g_{a\gamma}$ parameter space.}
    \label{fig:hostsuv} 
\end{figure*}

We first fit the observed energy spectrum using the ALP-only scenario (Figure~\ref{fig:OnlyALPs}), obtaining an AIC value of 85.86. For comparison, a fit based solely on the standard EBL model yields a substantially larger AIC of 192.46 and fails to reproduce the emission at both 18~TeV and 300~TeV. Although the ALP-only model successfully accounts for the 18 TeV emission, even the lower envelope of the fitted spectrum falls short of explaining the 300 TeV photon reported by Carpet-3. We then systematically evaluated the photon survival probability within the host galaxy, and Figure~\ref{fig:hostsuv} demonstrates that for both 18~TeV and 300~TeV photons, at least 70\% can escape from the host galaxy into intergalactic space across the explored parameter space of $m_a$ and $g_{a\gamma}$. Given that ALPs alone cannot explain the 300~TeV emission, and that a substantial fraction of very high energy photons can survive the host galaxy, LIV effects may play a crucial role in interpreting the extreme-energy emission of GRB~221009A.

Although LHAASO observations of GRB 221009A set stringent LIV limits ($E_{\rm QG,1} > 10^{20}$\,GeV, $E_{\rm QG,2} > 6.9 \times 10^{11}$\,GeV at 95\% CL) 
assuming no early time delay \citep{caoz2024}, the Carpet-3 detection of a 300\,TeV photon at \(T_0+4536\) s \citep{dzh2025} cannot be explained without new physics, where $T_0$ is the burst trigger time (Fermi-GBM onset). This photon should be absorbed by the EBL and arrives far too late for standard afterglow models. Ofengeim \& Piran \citep{ofe2025} demonstrate that only quadratic subluminal LIV simultaneously resolves both the EBL absorption and the late arrival puzzles \citep{son2025}, excluding linear ($n=1$) terms while favoring $\eta_2 = 1.30 \times 10^{-15}$ (95.4\% credibility), where $\eta_2$ is the dimensionless second-order LIV coefficient. We therefore adopt second-order subluminal LIV with $\eta_2 = 1.30 \times 10^{-15}$ as our physical benchmark.

\begin{figure}[!ht]
    \centering
    \includegraphics[width=5.0in, angle=0]{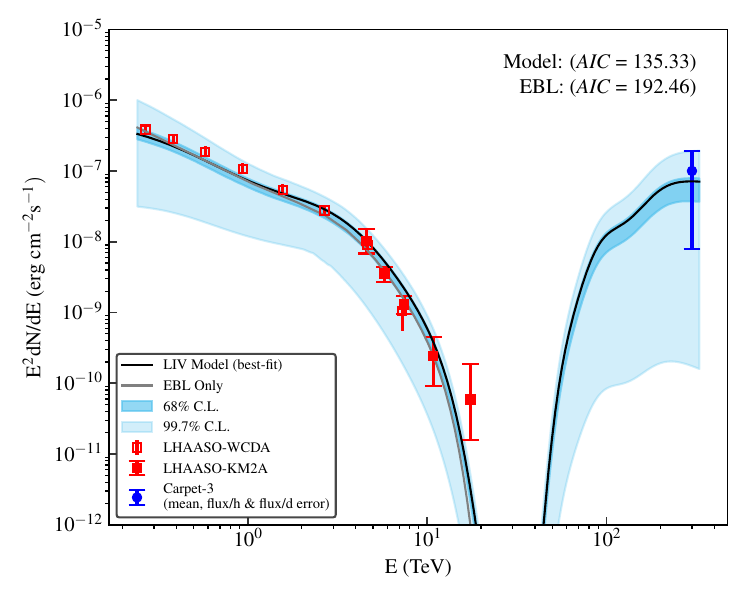}  
    \caption{Corresponding best-fit SEDs of GRB 221009A under only LIV scenario.}
    \label{fig:OnlyLIV} 
\end{figure}

We also attempted to fit the energy spectrum of GRB~221009A using the previously constrained LIV parameters, without considering the effect of ALPs. As shown in Figure~\ref{fig:OnlyLIV}, the fitted spectrum for the 300~TeV photon lies slightly below the observed average value and the value of AIC is 135.33, while the 18~TeV photon cannot be satisfactorily reproduced when accounting only for LIV.

The ALP-only model, while mitigating absorption at ~18 TeV, cannot produce sufficient flux at ~300 TeV because its oscillatory effect does not sufficiently elevate the pair-production threshold. Conversely, the LIV-only model, which raises this threshold, can allow the 300 TeV photon but tends to overproduce flux in the 10-30 TeV range compared to LHAASO data, as it lacks the energy-dependent modulation introduced by ALP mixing. The hybrid model naturally resolves this: ALP oscillations provide the necessary spectral shaping in the LHAASO band, while the LIV-induced threshold shift is crucial for enabling the survival of the Carpet-3 event.

\begin{figure}[!ht]
    \centering
    \includegraphics[width=5.0in, angle=0]{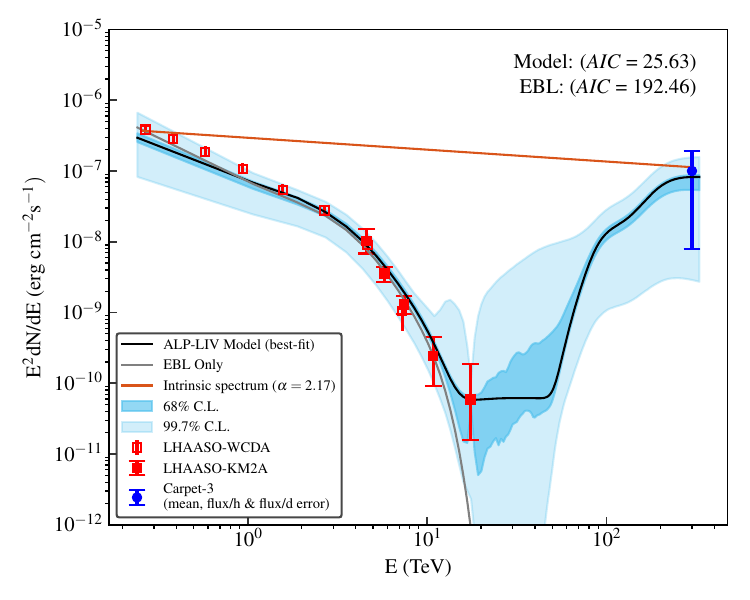}  
    \caption{Corresponding best-fit SEDs of GRB 221009A under ALP-LIV hybrid scenario with a coupling constant $ g_{a\gamma} = 1.685 \times 10^{-10} \, \mathrm{GeV}^{-1} $, $m_a = 9.545 \times 10^{-8} \, \mathrm{eV} $.}
    \label{fig:Mean} 
\end{figure}

We next apply the proposed ALP--LIV hybrid model to the TeV-band spectrum of GRB~221009A. For the LIV component, we adopt the quadratic ($n=2$) energy scale $E_{\rm LIV,2} = 1.30 \times 10^{-7} E_{\rm Pl}$ suggested by \citet{ofe2025}, while the ALP parameters are treated as free and optimized through spectral fitting. The best-fit values are $g_{a\gamma} = 1.685 \times 10^{-10}\,\mathrm{GeV}^{-1}$ and $m_a = 9.545 \times 10^{-8}\,\mathrm{eV}$. As illustrated in Figure~\ref{fig:Mean}, the hybrid model simultaneously accommodates both the 18~TeV emission and the $\sim$300~TeV photon within a unified framework. It yields the lowest Akaike Information Criterion value ($\mathrm{AIC} = 25.63$) among the models considered, indicating a statistically preferred description of the data.

\section{Discussion and Conclusions}\label{dis}

We have proposed and applied a unified framework that simultaneously incorporates photon--ALP oscillations (via the full evolution of the mixing matrix $M$, Eq.~(\ref{eq:MM})) and subluminal quadratic LIV effects (manifesting systematically in the average free path and threshold modifications embedded in $M$). This hybrid ALP-LIV approach provides greater flexibility and parameter freedom compared to treating either mechanism independently, as the LIV-induced suppression of pair production effectively modulates the effective optical depth over which ALP mixing occurs. We further simplify this unified model and apply the reduced formulation to interpret the multi-band spectral energy distribution of GRB 221009A. The analysis includes LHAASO observations (KM2A and WCDA arrays) during the 300-900~s interval after the trigger, as well as the 300~TeV photon reported by Carpet-3. Since differential flux values at 300~TeV are not explicitly provided in the Carpet-3 report~\citep{dzh2025}, we estimated its upper and lower limits, as well as an average value, by adopting one-hour and one-day time intervals for temporal differentiation. This procedure is consistent with the poorly constrained burst duration, ranging from roughly one hour to one day, and with the method of inferring flux by dividing the reported differential fluence by the chosen time window.

It is important that, in our model, LIV scenarios will introduce a preferred reference frame, commonly identified with the CMB comoving frame, photon–ALP mixing is naturally formulated in the local frame defined by the magnetic field. The relative velocity between these frames is of order ($v_{\rm pec}\sim 10^{-3}c$), as inferred from the CMB dipole. Corrections induced by boosting LIV-modified dispersion relations to the local frame are therefore suppressed by this small factor and remain subdominant with respect to the already highly suppressed LIV terms themselves. Consequently, the use of a common local propagation frame does not affect our results \citep[e.g.,][]{mat2005}. Besides, We only account for LIV effects exclusively via the $\gamma\gamma$ pair-production threshold modification. Although a complete LIV treatment would include changes to interaction amplitudes, cross sections, and additional vacuum processes, these effects are subdominant in the TeV–PeV regime. The threshold shift dominates the suppression of optical depth ($\tau_{\gamma\gamma}$) and therefore captures the leading LIV contribution to gamma-ray transparency.

More generally, several deliberate simplifications are adopted in this work to construct a tractable and physically transparent first-order framework. In particular, photon-ALP mixing in the IGMF is neglected. This approximation is well justified for weak intergalactic fields ($B_{\rm IGMF} \lesssim 1$~nG) and for the redshift of GRB~221009A, where plasma effects dominate at the highest energies considered and suppress efficient photon--ALP conversion. We note, however, that in scenarios involving stronger or more structured IGMFs, additional conversion could occur, potentially shifting the preferred ALP parameter space.

Likewise, we deliberately adopt only the EBL model of \citet{fin2022}. This model represents one of the most up-to-date EBL frameworks and is tightly constrained by a broad range of observational inputs, including luminosity densities, stellar mass functions, dust attenuation, and $\gamma$-ray absorption measurements, making it particularly well suited for the energy range ($\gtrsim 10$~TeV) explored here. Comparisons with other widely used EBL models (e.g., \citealt{sal2021}; \citealt{gil2012}) indicate that differences in the predicted optical depths above 10~TeV are typically within $\sim$30\%, well below the dominant systematic uncertainties associated with current VHE observations. Consequently, restricting our analysis to a single, well-motivated EBL model does not compromise the robustness of our main conclusions.

We further adopt representative average parameters for the magnetic fields and electron densities in both the GRB host galaxy and the Milky Way, following standard practice in the literature. While line-of-sight variations and environmental inhomogeneities could affect the precise photon-ALP conversion probabilities, they are unlikely to qualitatively alter the key conclusion that a high photon survival fraction is achievable. For LIV, we focus on the dominant modification of the pair-production threshold, which captures the leading effect relevant for enhanced transparency at ultra-high energies. A more complete effective-field-theory treatment could introduce subdominant spectral corrections, but these are not expected to overturn our main results. Taken together, these modeling choices demonstrate that the success of the ALP–LIV hybrid scenario does not stem from over-parameterization, but rather from its underlying physical mechanisms.

Finally, we note that VHE gamma rays propagating through the Milky Way may experience additional attenuation via electron-positron pair production on the interstellar radiation field (ISRF), whose optical depth depends sensitively on the line of sight \citep{por2017,dim2025}. For GRB~221009A, which is located at relatively high Galactic latitude, ISRF-induced absorption is expected to be weak and does not qualitatively affect our conclusions within the simplified framework adopted here. Future studies could incorporate line-of-sight-dependent Galactic models, such as detailed electron density and radiation-field distributions \citep[e.g.,][]{yao2017}, to further assess the impact of these environmental uncertainties on photon survival probabilities.

The explored ALP parameter space spans $\log_{10}(m_a / 10^{-8}\,\mathrm{eV}) \in [-2,\,1]$ and $\log_{10}(g_{a\gamma} / \mathrm{GeV}^{-1}) \in [-12,\,-9]$, consistent with current experimental and astrophysical bounds. For the LIV component, we fix the quadratic ($n=2$) coefficient to $\eta_2 = 1.30 \times 10^{-15}$, corresponding to an energy scale $E_{\rm LIV,2} = 1.30 \times 10^{-7} E_{\rm Pl}$, as suggested by recent analyses of the Carpet-3 $\sim$300~TeV photon \citep{ofe2025}. That study favors a quadratic subluminal modification at the 95.4\% credibility level while disfavoring linear ($n=1$) LIV contributions. Independent constraints derived solely from LHAASO observations of GRB~221009A yield limits of $E_{\rm QG,2} \gtrsim (6$--$10)\times10^{11}$~GeV \citep{caoz2024}, corresponding to $|\eta_2| \lesssim 10^{-14}$~-~$10^{-15}$ depending on the specific formulation. Our adopted value therefore lies comfortably within, or close to, the allowed parameter space. Tighter exclusions reported from other GRBs or blazar observations mainly constrain linear or superluminal LIV scenarios, leaving quadratic subluminal effects at this level viable. For the ALP component, the coupling $g_{a\gamma}$ and mass $m_a$ are treated as free parameters and optimized through spectral fitting. As shown in Figure~\ref{fig:hostsuv}, across the explored $m_a$–$g_{a\gamma}$ parameter space, at least $\sim$70\% of the 18~TeV and 300~TeV photons are able to escape the host galaxy. This high survival fraction implies that LIV-induced modifications remain significant even with efficient photon-ALP conversion.

As shown in Figures~\ref{fig:OnlyALPs}, \ref{fig:OnlyLIV}, and \ref{fig:Mean}, the hybrid ALP--LIV model successfully reproduces both the 18~TeV emission detected by LHAASO and the $\sim$300~TeV photon reported by Carpet-3, yielding the lowest AIC value of 25.63 among all considered models. In the ALP-only scenario, the best-fit parameters are $g_{a\gamma} = 1.489 \times 10^{-10},\mathrm{GeV}^{-1}$ and $m_a = 4.191 \times 10^{-7},\mathrm{eV}$; however, this model fails to reproduce even the lower bound of the observed 300~TeV flux, resulting in a significantly larger AIC of 85.86. Conversely, the LIV-only scenario with quadratic ($n=2$) dispersion modifications slightly underestimates the 300~TeV flux and cannot adequately account for the 18~TeV emission, yielding an AIC of 135.33. Only the combined ALP--LIV framework, with best-fit parameters $g_{a\gamma} = 1.685 \times 10^{-10},\mathrm{GeV}^{-1}$ and $m_a = 9.545 \times 10^{-8},\mathrm{eV}$, provides an excellent and self-consistent description of all available observational data. In addition, we also performed spectral fitting to the multi-wavelength data of GRB~221009A using a standard EBL absorption model alone. As shown in Fig.~1,3,4, the resulting spectrum fails to account for the emission at both 18~TeV and $\sim$300~TeV, and yields a substantially larger Akaike information criterion value, AIC=192.46, indicating a poor description of the data.

A key issue is whether the inferred parameters are compatible with independent constraints from other experiments and observations. For the ALP sector, a coupling $g_{a\gamma} \sim 10^{-10},\mathrm{GeV}^{-1}$ at $m_a \sim 10^{-7},\mathrm{eV}$ lies close to the upper envelope of current limits but is not formally excluded. Searches for spectral irregularities in blazar samples with \emph{Fermi}-LAT \citep[e.g.,][]{aje2016}, as well as stacked analyses of Galactic and extragalactic sources, typically constrain $g_{a\gamma} \gtrsim (0.5$--$1)\times10^{-10},\mathrm{GeV}^{-1}$ in the neV–$\mu$eV mass range at 95\% confidence, with complementary bounds from helioscope experiments such as CAST at lower masses. These limits generally assume dominant ALP–photon mixing in Galactic or intergalactic magnetic fields and rely on null detections of oscillatory features. In contrast, our scenario exploits specific line-of-sight magnetic-field configurations together with an additional LIV-induced enhancement of transparency, which can partially relax constraints derived under pure-ALP assumptions. As a result, no direct inconsistency arises: the favored parameters are marginally compatible with existing bounds or reside in comparatively less constrained regions once the LIV contribution is included.

In our fits, the ALP parameters primarily affect the spectral shape in the 10–30 TeV range, while LIV dominates above 100 TeV. Concerning possible degeneracies between ALP and LIV parameters, we find that the two mechanisms are not strongly degenerate within the parameter space explored here. ALP-photon mixing, which is oscillatory and magnetic-field dependent, primarily affects photon survival through conversion and reconversion processes, whereas LIV acts by systematically modifying the $\gamma\gamma$ pair-production threshold and, consequently, the effective propagation encoded in the mixing matrix $M$. These effects are therefore complementary. ALPs can induce spectral hardening or modulations over specific energy ranges, while quadratic subluminal LIV produces a monotonic increase in transparency at the highest energies ($E \gtrsim 10$–$100$~TeV), enabling a natural explanation of the Carpet-3 $\sim$300~TeV event without overproducing the LHAASO flux at lower energies. Fits to the combined LHAASO and Carpet-3 spectrum confirm that ALP-only models struggle to accommodate the ultra-high-energy data, LIV-only models yield only modest improvements in the TeV regime, and the hybrid ALP-LIV scenario provides a significantly improved description across the full 10-300~TeV range.

We note, however, that degeneracies could emerge in other regions of parameter space. For example, stronger ALP couplings combined with weaker LIV effects (or vice versa) might reproduce similar photon survival probabilities at specific energies or redshifts. Such degeneracies are more likely for smaller $g_{a\gamma}$ or larger $m_a$, where ALP effects diminish, or for alternative LIV orders and energy scales. A comprehensive exploration of the joint $(g_{a\gamma}, m_a, \eta_2)$ parameter space would be required to fully quantify these possibilities, ideally informed by future observations with CTA, or next-generation GRB detections. Within the parameter region favored by GRB~221009A, however, we find no evidence for strong degeneracy, supporting the interpretation that the observed extreme-energy transparency arises from a genuine synergistic interplay between ALP mixing and LIV effects.

This work develops a unified hybrid framework that combines ALP oscillations with LIV to account for the anomalous transparency of the Universe to ultra-high-energy gamma rays from GRB~221009A. In contrast to previous studies that typically considered ALP and LIV effects separately, such as \citet{gal2025}, who focused primarily on pure LIV-induced threshold modifications and associated time delays, however, we perform a joint Bayesian analysis of the time-resolved TeV spectrum, incorporating both the LHAASO data in the 300-900~s interval and the $\sim$300~TeV Carpet-3 event. Our results quantitatively show that neither ALP nor LIV alone can simultaneously reproduce the observed 18~TeV and 300~TeV photons, whereas their combination provides a statistically preferred description of the data, as quantified by the AIC.

In addition, our result are broadly consistent with those of \citet{sat2025}, who also investigated an ALP+LIV scenario for GRB~221009A and found that a joint mechanism is required to explain the observed spectrum. Key differences lie in our use of time-resolved differential fluxes rather than integrated fluence, a more careful treatment of the flux uncertainty associated with the 300~TeV photon through both one-hour and one-day temporal windows, and the adoption of the quadratic subluminal LIV benchmark motivated by \citet{ofe2025}. These methodological choices allow for a more detailed characterization of the afterglow-phase emission and the sensitivity of the inferred parameters. While some aspects of the ALP–LIV implementation in \citet{sat2025} are formulated differently, their analysis provides valuable complementary support for the viability of the hybrid scenario. Finally, We emphasize that our work is not intended to be a fine-tuned solution tailored solely to the high-energy spectrum of GRB~221009A. Rather, we aim to construct a self-consistent and physically motivated ALP-LIV hybrid framework, which can be applied more generally to make testable predictions for the propagation and detection of gamma-ray photons with energies $\gtrsim 10$~TeV in future observations.

Taken together, our analysis demonstrates that the anomalous transparency of GRB~221009A at energies extending from $\sim$10~TeV up to the $\sim$300~TeV regime cannot be accommodated by standard EBL absorption, nor by ALP or LIV effects acting in isolation. Instead, a hybrid ALP--LIV framework provides a unified and physically consistent explanation, in which photon-ALP oscillations and quadratic subluminal LIV act in complementary ways to enhance photon survival over cosmological distances. In this framework, the observed 18~TeV and $\sim$300~TeV events are simultaneously reproduced with the best overall statistical performance among the tested scenarios, while standard EBL-only, ALP-only, or LIV-only models fail to do so. By adopting literature-motivated LIV parameters and simplified but well-justified astrophysical environments, we show that this agreement arises from the core physical mechanisms of the model rather than from excessive freedom in parameter choice, establishing the ALP-LIV synergy as a viable interpretation of the GRB~221009A observations.

Looking ahead, the enhanced sensitivity of the Cherenkov Telescope Array (CTA) will be crucial for detecting a larger sample of $>10$~TeV photons from GRBs and AGN, enabling detailed spectral studies capable of distinguishing oscillatory transparency features characteristic of ALP-induced effects from the monotonic behavior expected in LIV scenarios. Independent constraints on LIV from future GRB time-delay analyses \citep[e.g.,][]{caoz2024}, together with dedicated ALP searches by next-generation experiments such as the International Axion Observatory (IAXO; \citet{ira2011}), will probe the ALP coupling--mass parameter space down to $g_{a\gamma} \sim 10^{-12}$--$10^{-10}\,\mathrm{GeV}^{-1}$ for $m_a \sim 10^{-3}$--$10^{-2}\,\mathrm{eV}$. This region partially overlaps with, yet remains complementary to, the higher-mass and lower-coupling parameter space explored by very-high-energy gamma-ray spectral modeling, including the present work. Such complementarity will enable progressively more stringent tests of the hybrid ALP--LIV framework. In parallel, implementing LIV-modified absorption self-consistently within public gamma-ray propagation frameworks (e.g., through extensions of \texttt{gammaALPs}; \citet{mey2022b}) will allow this hybrid scenario to be applied to source populations, enabling robust statistical tests. More comprehensive implementations within advanced propagation codes such as \texttt{CRPropa}, incorporating existing modules for LIV \citep{sav2024} and ALP effects \citep{alv2023}, would further enable detailed Monte Carlo simulations of electromagnetic cascades and population-level studies. The detection of GRB~221009A thus highlights a possible interplay between particle physics beyond the Standard Model and quantum-gravity effects. Our hybrid ALP--LIV framework provides a concrete, testable paradigm for this synergy, making specific predictions for the very-high-energy sky that forthcoming observatories will be well positioned to probe.

\acknowledgments
The authors would like to thank the anonymous referee for the helpful comments, which improved the manuscript significantly. The authors gratefully acknowledge the financial supports from the National Natural Science Foundation of China (grants 12063005, 12063006, 12233006, U2031111), the program for Innovative Research Team (in Science and Technology) in University of Yunnan Province (IRTSTYN), the program for Reserve Talents of Young, Middle-aged Academic and Technical Leaders in Yunnan Province (grant 202205AC160087, 202405AC350114), the program for Hunan Outstanding Youth Science Foundation (2024JJ2040) and the Special Basic Cooperative Research Programs of Yunnan Provincial Undergraduate Universities' Association (grants NO.202501BA070001-122). The authors gratefully acknowledge the computing support provided by the JRT Science Data Center at Yuxi Normal University and Q.L.H. gratefully acknowledges the kind and selfless help from Prof. Nikita Pozdnukhov \citep{dzh2025} and Prof. Dom{\'\i}nguez, A.\citep{dom2013}.
%


\clearpage

\end{document}